\documentclass[manuscript]{aastex62}

\hypersetup{breaklinks}   

\received{\today}
\submitjournal{ApJ}

\shorttitle{Heating of the solar chromosphere by acoustic waves}
\shortauthors{B.~Ku\'zma et al.}
\newcommand{\beqa}{\begin{eqnarray}}
\newcommand{\eeqa}{\end{eqnarray}}

\usepackage{amssymb}
\usepackage{amsmath}
\usepackage[varg]{txfonts}
\usepackage{natbib}
\usepackage{url}             
\usepackage{graphicx}
\usepackage{xcolor}
\usepackage{float}
\newcommand{\beq}{\begin{equation}}
\newcommand{\eeq}{\end{equation}}
\bibpunct{(}{)}{;}{a}{}{,} 

\usepackage{textcomp}
\usepackage{wrapfig}
\newcommand{\foo}{\rlap{+}{\texttimes}}

\begin{document}




\title{\bf \large Heating of a quiet region of the solar chromosphere \\
by ion and neutral acoustic waves}

\author{B. Ku\'zma}
\affiliation{Group of Astrophysics, Institute of Physics, University of M. Curie-Sk{\l}odowska, \\
ul. Radziszewskiego 10, 20-031 Lublin, Poland}

\author{D. W\'ojcik}
\affil{Group of Astrophysics, Institute of Physics, University of M. Curie-Sk{\l}odowska, \\
ul. Radziszewskiego 10, 20-031 Lublin, Poland}

\author{K. Murawski}
\affiliation{Group of Astrophysics, Institute of Physics, University of M. Curie-Sk{\l}odowska, \\
ul. Radziszewskiego 10, 20-031 Lublin, Poland}





%
\begin{abstract}
Using high-resolution numerical simulations we investigate the plasma heating driven by periodic two-fluid 
acoustic waves that originate at the bottom of the photosphere and propagate into the gravitationally stratified and partially ionized solar atmosphere.
We consider ions+electrons and neutrals as separate fluids that interact between themselves via collision forces. 
The latter play an important role in the chromosphere, leading to significant damping of short-period 
waves. 
Long-period waves do not essentially alter the photospheric temperatures, but they exhibit the capability of depositing a part of their energy in the chromosphere. This results in up about a five times increase of ion temperature that takes place there on a time-scale of a few minutes. The most effective heating corresponds to waveperiods within the range of about 30-200~s with a peak value located at 80 s. 
However, we conclude that for the amplitude of the driver chosen to be equal to 0.1 km s$^{-1}$, this heating is too low to balance the radiative losses in the 
chromosphere. 
\end{abstract}





%
\section{Introduction}
One of the main open questions of solar physics is energy transport from the photosphere to the chromosphere 
and further up to the corona. 
It is well known that chromospheric plasma radiates more than the corona, 
and thus the additional source of chromospheric heating is required \citep[e.g.,][]{Narain1996}. The photosphere is full-filled by a wide range of oscillations and it is considered as a main source of waves in the solar atmosphere. Among others, the excited acoustic waves propagate into upper, quiet and essentially magnetic-free atmospheric regions \citep[e.g.,][]{Nakariakov2005}. So far the wave processes were investigated mainly within the framework of magnetohydrodynamics  
which is a good model for a fully-ionized plasma. However, the photospheric and chromospheric plasma is only partially-ionized 
and neutrals play an important role in their dynamics \citep[e.g.,][]{Zaqarashvili2011}. 
The ion-neutral collisions may lead to diversity of phenomena such as 
for instance wave damping \citep[e.g.,][]{Khodachenko2004,Forteza2007,Piddington1956,Watanabe1961,Kulsrud1969,Haerendel1992,DePontieu1998,James2002,Erdelyi2004}.




The idea that acoustic waves can heat the solar atmosphere was devoted by \cite{Biermann1946} and \cite{Schwarzschild1948} who proposed for the first time that 
acoustic waves can be recognized as the main heating agents of the chromosphere with wavefront steepening with height and dissipation in shocks in the chromosphere being considered 
as the main mechanism behind plasma heating. 
Later on, the problem of chromospheric heating by acoustic waves was investigated widely. 
For instance, \cite{Carlsson1995} questioned plasma average temperature growth in the acoustically heated chromosphere. 
\cite{Fossum2005, Fossum2006} performed numerical simulations of high-frequency acoustic waves 
and concluded that they are not sufficient to heat the chromosphere. 
On the other hand, \cite{Ulmschneider2003} stated that acoustic waves are main source of heating 
in non-magnetic regions of the chromosphere and \cite{Cuntz2007} provided significant evidence 
that this heating can be locally dominant. 
Using two-fluid numerical simulations \cite{Maneva2017} showed that magnetoacoustic waves, 
while propagating through gravitationally stratified medium of the chromosphere, 
lead to alterations of plasma temperature. 


We aim to contribute to the investigations mentioned above by performing novel and high-resolution two-fluid simulations of acoustic waves propagating in the solar atmosphere and to quantify amount of heating occurring in the chromosphere. 
This paper is organized as follows. 
In Section 2, we describe the physical model of the solar atmosphere. 
Numerical simulations are presented in Section 3. 
Summary and conclusions are outlined in the last section.
\section{Physical model}
In this part of the paper we describe the physical model of the solar atmosphere we employed. 
%
%
%
In particular, 
we consider a gravitationally stratified and partially-ionized solar atmosphere. 
With use of a realistic height-dependent temperature profile of 
the semi-empirical model of \cite{Avrett2008}  
we determine uniquely the equilibrium mass density and gas pressure profiles 
which fall off with height. 
To describe chromospheric plasma we adopt the set of equations for two fluids, 
mainly for ions+electrons and neutrals, with contribution of both species depending on local ionization level. 
These two-fluid equations were derived by a number of authors 
\citep[e.g.,][and references therein]{Smith2008,Zaqarashvili2011,Meier2012,Soler2013,Ballester2018}. 
They were implemented into the JOANNA code \citep{Wojcik2017} 
which was recently used by \cite{Kuzma2017} and \cite{Srivastava2018} to simulate two-fluid jets and by \cite{Wojcik2018a} to study two-fluid acoustic cut-off periods. 
In our model we assume initial thermal balance between ion and neutral components of plasma, 
$T_{\rm i}=T_{\rm n}=T_{0}$ (Oliver et al. 2016) 
and neglect all electron-components due to the small mass of electrons in relation to ions and neutrals. Note that the subscripts $_{\rm i}$ and $_{\rm n}$ correspond to ions (protons) and neutrals (hydrogen atoms), respectively. See also Section 3 of \cite{Wojcik2018a} for extended discussion on two-fluid hydrostatic equilibrium. 

%

The two-fluid equations are taken from \cite{Zaqarashvili2011} with exception for the heat production terms in the total energy equations which result from ion-neutral collisions. These terms are written here in similar (but generalized here for the solar mean atomic masses) form as in \cite{Oliver2016}. See also \cite{Wojcik2018a}.  
Thus ion and neutral components of plasma are governed by the following set of equations: 
%
\beqa
\label{eq:MHD_rho}
&{{\partial \varrho_{\rm i}}\over {\partial t}}+\nabla \cdot (\varrho_{\rm i}{{\bf V} _{\rm i}})=0\, ,\\
\label{eq:MHD_V}
&{{\partial \varrho_{\rm n}}\over {\partial t}}+\nabla \cdot (\varrho_{\rm n}{{\bf V} _{\rm n}})=0\, ,\\
\label{eq:MHD_V2}
&\varrho_{\rm i}\left[\frac{\partial{\bf V_{\rm i}}}{\partial t} +({\bf V}_{\rm i}\cdot \nabla){\bf V_{\rm i}} \right]=-\nabla p_{\rm i \, e}+\varrho_{\rm i}{\bf g} - 
\alpha_{\rm in}({\bf V}_{\rm i}-{\bf V}_{\rm n})\, , \\
\label{eq:MHD_p}
&\varrho_{\rm n}\left[\frac{\partial{\bf V_{\rm n}}}{\partial t} +({\bf V}_{\rm n}\cdot \nabla){\bf V_{\rm n}} \right]=-\nabla p_{\rm n} +\varrho_{\rm n}{\bf g} +\alpha_{\rm in}({\bf V}_{\rm i}-{\bf V}_{\rm n})\, , \\
\label{eq:MHD_p2}
&\frac{\partial E{\rm {\rm _i}}}{\partial t}+\nabla\cdot((E{\rm {\rm _i}}+p_{\rm {\rm ie}}){\bf V_{\rm i}}) = -\alpha_{\rm in}{\bf V}_{\rm i}({\bf V}_{\rm i}-{\bf V}_{\rm n}) + Q^{\rm in}_{\rm i} +\varrho{\rm {\rm _i}} {\bf g} \cdot {\bf V_{\rm i}}\,,\\
\label{eq:MHD_p}
&\frac{\partial E_{\rm n}}{\partial t}+\nabla\cdot((E_{\rm n}+p_{\rm n}){\bf V_{\rm n}}) = \alpha_{\rm in}{\bf V}_{\rm n}({\bf V}_{\rm i}-{\bf V}_{\rm n}) + Q^{\rm in}_{\rm n} +\varrho_{\rm n} {\bf g} \cdot {\bf V_{\rm n}}\,,\\
\label{eq:MHD_p2}
&Q^{\rm in}_{\rm i}=\alpha_c \left[\frac{1}{2} |{\mathbf V}_{\rm i}-{\mathbf V}_{\rm n} |^2 + \frac{3k_{\rm B}}{m_{\rm H}(\mu_{\rm i}+\mu_{\rm n})}\left(T_{\rm n}-T_{\rm i} \right )\right] \, , \\
\label{eq:MHD_p2}
&Q^{\rm in}_{\rm n}=\alpha_c \left[\frac{1}{2} |{\mathbf V}_{\rm i}-{\mathbf V}_{\rm n} |^2 + \frac{3k_{\rm B}}{m_{\rm H}(\mu_{\rm i}+\mu_{\rm n})}\left(T_{\rm i}-T_{\rm n} \right )\right] \, . 
\eeqa
These equations are supplemented by the ideal gas laws as 
\begin{equation}
p_{\rm n}=\frac{k_B}{m_{\rm H}\,\mu_{\rm n}}\varrho_{\rm n} T_{\rm n}\,,\hspace{3mm} p{\rm {\rm _i}} = \frac{2k_B}{m_{\rm H}\,\mu_{\rm i}}\varrho{\rm {\rm _i}} T{\rm {\rm _i}}\, .
\end{equation}
Here $E_{\rm i}$ and $E_{\rm n}$ are respectively ion and neutral energy densities, 
\begin{equation}
E_{\rm i} = \frac{p_{\rm {\rm i e}}}{\gamma-1} + \frac{1}{2}\varrho{\rm {\rm _i}} {\bf V{\rm _i}}^2 \, , \hspace{3mm} E_{\rm n} = \frac{p_{\rm n}}{\gamma-1} + \frac{1}{2}\varrho_n {\bf V_{\rm n}}^2\, ,
\end{equation}
$\gamma=5/3$ is the ratio of specific heats, $\alpha_{in}$ and $\alpha_{ni}$ are friction coefficients between ions and neutrals, $\mu_{\rm i}$ and $\mu_{\rm n}$ are mean atomic masses of respectively ions and neutrals, which are taken from the solar abundance model \citep[e.g.,][]{Vogler2005}, $t$ is time, $k_{B}$ is Boltzmann constant, $m_{H}$ hydrogen mass, $T_{\rm i}$ and $T_{\rm n}$ temperatures of respectively ions and neutrals, ${\bf V}_{\rm i}$ and ${\bf V}_{\rm n}$ their corresponding velocities, and $p_{\rm i}$ and $p_{\rm n}$ are gas pressures. A constant solar gravity of its magnitude g=274.78 m s$^{-1}$ points in the negative $y-$direction. To estimate $\alpha_{\rm in}$ we use formula provided by \cite{Braginskii1965} 
with the collisional cross-section taken as quantum-mechanical from \cite{Vranjes2013} who showed that the classical hard-sphere model may lead to underestimation of the cross-section values, and they derived from quantum-mechanical model the integral cross section $\sigma_{\rm in}$ for ions (protons) collisions with neutrals (neutral hydrogen atoms). For typical chromospheric plasma temperature in the range of $6-10\cdot 10^3$ K this cross-section is equal to $1.89\cdot 10^{-18}$ m$^{2}$, that is about three orders of magnitude larger than in the hard-sphere model. 
Following \cite{Zaqarashvili2011}, we assumed $\alpha_{in}=\alpha_{ni}$. However, ion-neutral collision frequency differs from neutral-ion collision frequency \citep{Ballester2018}. 
\begin{figure}[!ht]
       \begin{center}
        \mbox{
           \hspace{-0.5cm}
                \includegraphics[scale=0.65, angle=0]{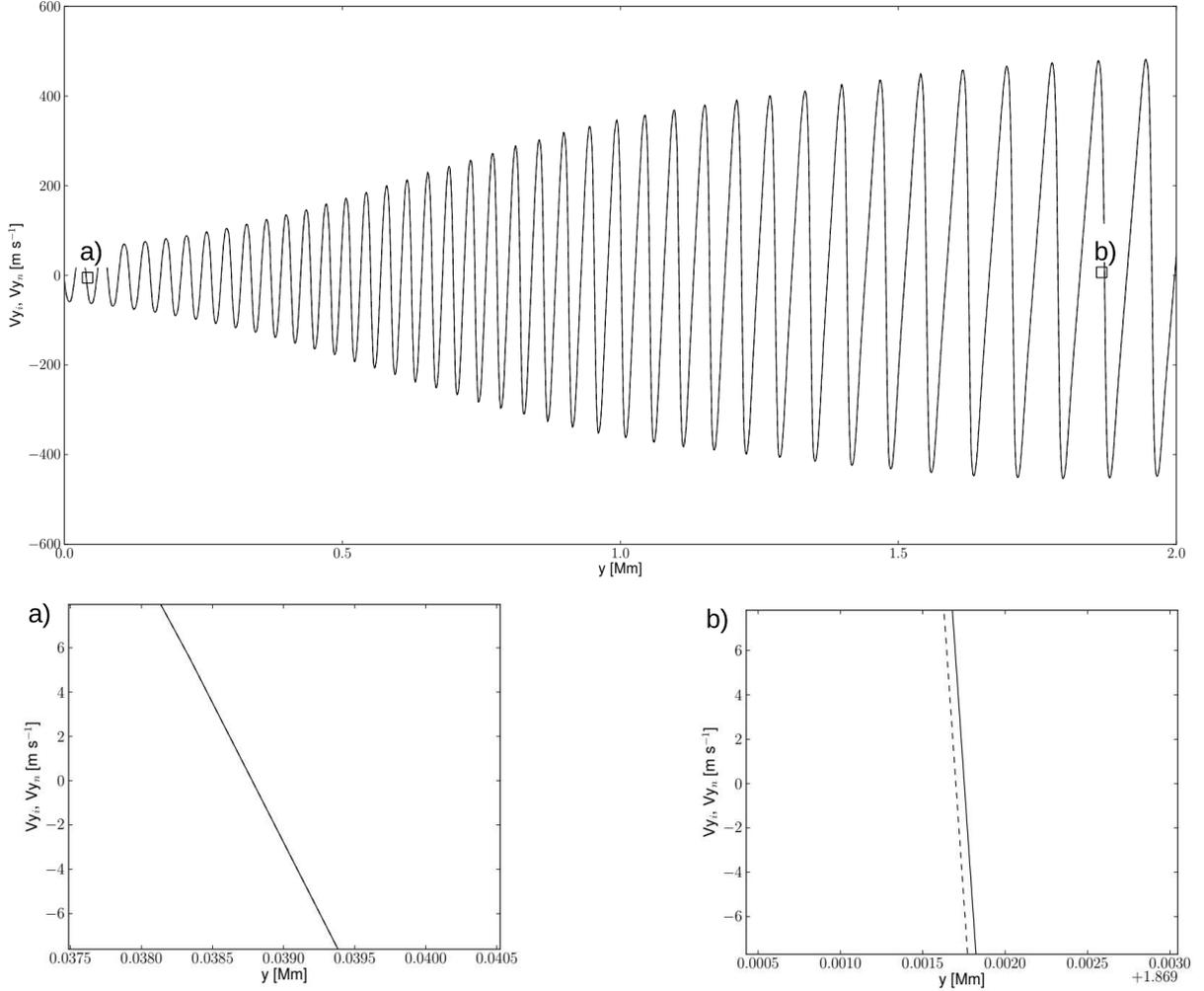} 
                }
                \caption{\small Vertical components of ion, $V_{{\rm i} \, y}$, (solid) and neutral, $V_{{\rm n} \, y}$, (dashed) velocities 
                vs height, $y$, 
                drawn at time $t=10^4$ s for $P_{\rm d}=5$ s. The bottom panels show zoomed in regions a) and b).  
                        }
                \label{fig:1}
                        \end{center}
\end{figure}
At the bottom of the photosphere, given by $y=0$, we set the periodic driver in vertical component of ion and neutral velocities, i.e. 
\begin{equation}
V_{{\rm i,n} \, y}(y,t)=V_{\rm 0} \, \sin{\left(\frac{2\pi t}{P_{\rm d}}\right)} \, ,
\end{equation}
where $V_{\rm 0}$ is the amplitude of the driver 
and $P_{\rm d}$ its period. 
This driver excites upwardly propagating ion and neutral acoustic waves \citep[e.g.,][]{Zaqarashvili2011}.  
We set $V_{0}=0.1$ km s$^{-1}$, 
but allow 
$P_{\rm d}$ 
to vary within the range of 
$5$~s $ \leq P_{\rm d} \leq 300$~s. 
These values fit to the typical characteristics of flow associated with 
the solar granulation \citep[e.g.,][]{Musielak1994,Hirzberger2003}. 
%
Among others \cite{Tu2013} and \cite{Arber2016} determined amplitude of Alfv\'en waves with use of piecewise power law. Thus, more realistic ways of implementing the amplitude $V_0$ are possible. 

\begin{figure}[!ht]
       \begin{center}
       \mbox{
                  \hspace{-0.5cm}
                \includegraphics[scale=0.45, angle=0]{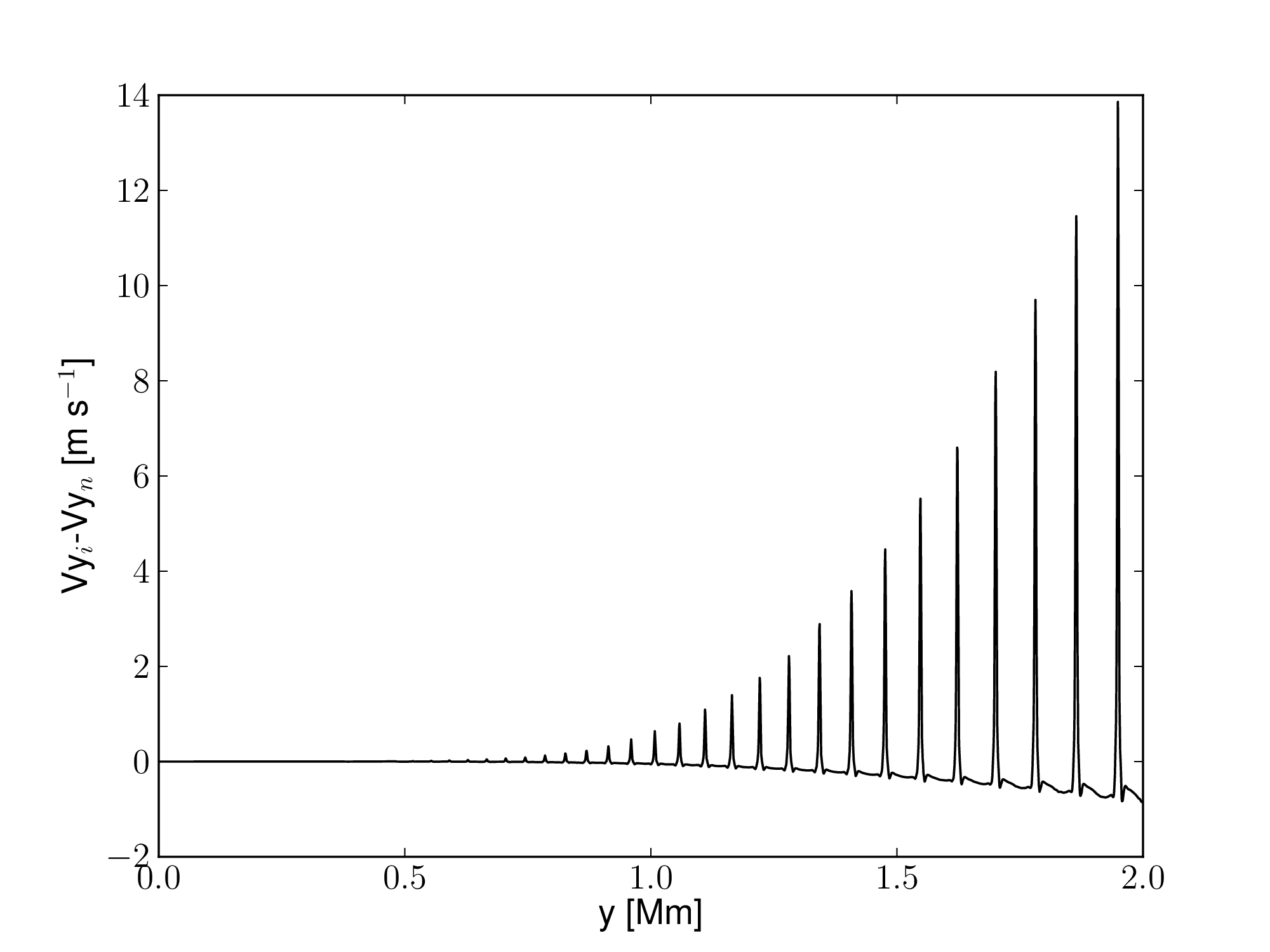}      
                }         
                \caption{\small Difference between vertical components of ion ($V_{{\rm i} \, y}$) and neutral ($V_{{\rm n} \, y}$) velocities vs height, $y$, at $t=10^4$~s for $P_{\rm d}=5$ s. 
                       }
                \label{fig:01}
        \end{center}
\end{figure}

\section{Numerical results}

We solve two-fluid system of equations numerically with use of the JOANNA code \citep{Wojcik2017}. 
In our problem, we set the Courant-Friedrichs-Lewy number \citep{Courant1928} equal to 0.8 
and adopt WENO3 with HLLD Riemann solver \citep{Miyoshi2010}. 
Typically, our one-dimensional (1D) numerical box covers the region between the bottom of the photosphere ($y=0$~Mm) 
and the low corona ($y=2.5$~Mm), and by default it is represented by $25\times 10^{3}$ identical numerical cells. 
This results in the uniform spatial grid of its size $\Delta y=100$ m. 
Above this region, namely for $2.5$~Mm $<y<$ $30$~Mm, we stretch the grid, 
dividing it into $128$ cells with their size growing with height. 
As short waveperiod waves 
require very fine spatial resolution 
we 
refine our spatial grid for shorter $P_{\rm d}$ to match better resolution. 


%

%

%
We investigate now impact of acoustic waves propagating through the photosphere and chromosphere on plasma heating. 
Figure \ref{fig:1} shows the vertical profile of $V_{{\rm i} \, y}$ (solid line) and $V_{{\rm n} \, y}$ (dashed line) 
drawn at $t=10^4$~s for $P_{\rm d}=5$~s (top) and the zoomed in regions in the photosphere (panel a) and upper chromosphere (panel b). 
In the photosphere, where ions and neutrals are strongly coupled both $V_{{\rm i} \, y}$ and $V_{{\rm n} \, y}$ essentially overlap each other (panel a). Higher up, in the chromosphere and upper chromosphere, where ions and neutrals start to decouple, the difference between ion and neutral wavefront positions is clearly seen. 
We infer that amplitudes of 
the excited ion 
and neutral 
acoustic waves 
are essentially not affected by ion-neutral collisions in the photosphere. 
However, higher up the amplitude grows e-times with the pressure-scale height, while waves steepen into shocks. 
\begin{figure*}[!ht]
        \begin{center}
        \mbox{
        \hspace{-0.5cm}
                \includegraphics[scale=0.65, angle=0]{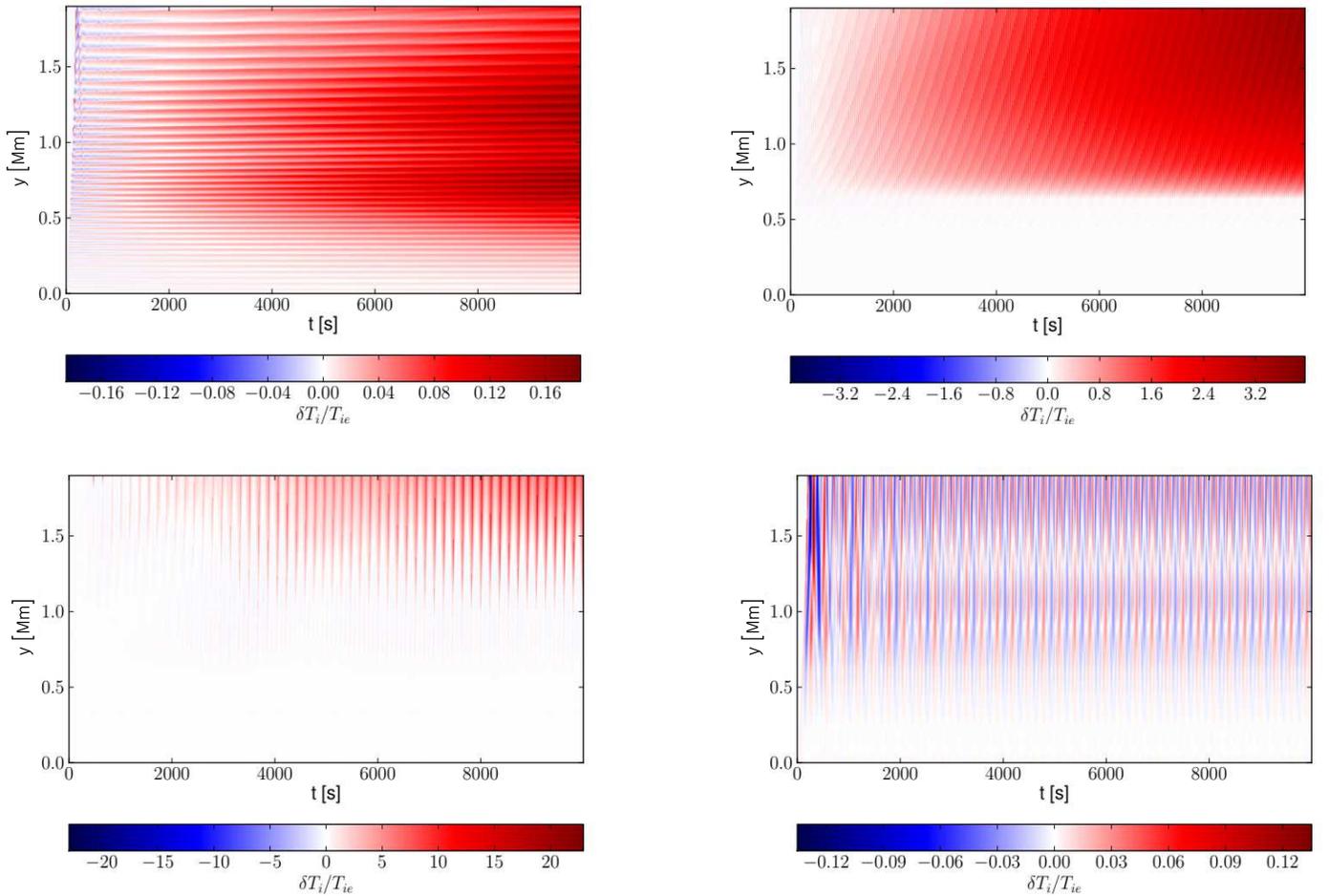}\hspace{-0.5cm}
                }
                      
                \caption{\small Time-distance plots of relative perturbed temperature of ions, $\delta T_{\rm i}/T_{\rm i}$, in the case of the driving period 
                $P_{\rm d}=5$~s, 
                $P_{\rm d}=30$~s, 
                $P_{\rm d}=180$~s, and $P_{\rm d}=300$~s (from top-left to bottom-right). 
                        }
                \label{fig:2}
        \end{center}
\end{figure*}

From Eqs.~(7) \& (8) we infer that the velocity difference between ions and neutrals, 
%
 $\delta V_{y}=V_{{\rm i} \, y}-V_{{\rm n} \, y}$, 
%
plays a role in plasma heating associated with propagating acoustic waves. 
Figure \ref{fig:01} illustrates $\delta V_{y}(y)$ 
for the 
driving period $P_{\rm d}=5$~s 
at $t=10^{4}$~s. 
At the bottom of the photosphere both ions and neutrals remain strongly coupled 
and therefore they propagate with essentially same velocity. 
The difference between ion and neutral velocities grows with height, 
reaching a magnitude of about 14~m~s$^{-1}$ at the transition region which is located at $y\approx 2.1$~Mm. As at higher altitudes the acoustic wave profiles steepen into shocks, these differences are present at these shocks and they result from structured nature of shocks in two-fluids regime \citep{Hillier2016}, which are well seen in Fig.~\ref{fig:1} (top and panel b).  


Figure~\ref{fig:2} shows the temporal and spatial evolution of relative perturbed temperature of ions, $\delta T_{\rm i}/T_{\rm i}=(T_{\rm i}-T_{0})/T_{0}$, where $T_{0}(y)$ is the hydrostatic temperature \citep{Avrett2008}. 
The low-period waves, mainly of $P_{\rm d}=5$~s, are significantly damped with height (top-left), 
and this damping results from ion-neutral collisions. 
It is expected, as for oscillation periods larger than ion-neutral collision times, 
the damping efficiency supposed to decrease and e-times growth of the wave amplitude with a pressure scale-height takes over a leading role. 
During the damping process the energy carried by shocking acoustic waves is dissipated in the upper photosphere and lower chromosphere.  
For $P_{\rm d}=5$ s the increase of the initial temperature is up to 16\% on time-scale of $10^3$~s, 
and the heating occurs mainly from the level $y=0.6$~Mm up to $y=0.8$~Mm and from $y=1.1$~Mm up to $y=1.4$~Mm.

According to the linear theory \citep[e.g.,][]{Zaqarashvili2011} acoustic waves with longer periods, and thus with their oscillation frequencies much lower than 
ion-neutral collision frequencies, are weakly damped and they possess capability to propagate upward, 
reaching upper atmospheric regions if their periods are lower than cut-off periods \citep{Wojcik2018}.  
Wave amplitude growth with height is a dominant factor over the wave amplitude damping resulting from ion-neutral collisions. As a result, the wave amplitude experiences a net growth with height and the linear theory is too approximate. 

The top-right panel of Fig.~\ref{fig:2} corresponds to $P_{\rm d}=30$~s. 
Acoustic waves deposit their energy mainly in the chromosphere, at the height of $y>0.5$~Mm. 
The waves propagate higher up. However, as the damping depends on ion-neutral collision coefficient (Eqs. 1-2), 
and thus effectively on the mass density and temperature, 
higher up the amplitude of disturbance is not sufficient to heat the chromosphere significantly. 
As $P_{\rm d} = 180$~s is lower than acoustic cutoff period \citep{Wojcik2018}, according to the theory originally developed by \cite{Lamb1909} 
the excited acoustic waves propagate freely across the chromosphere up to the transition region (bottom-left panel). See \cite{Wojcik2018a} for the corresponding numerical simulations. 
For $P_{\rm d}=300$ s cooling events discernible in Fig. 4 are related to the fluctuations of perturbed ion temperature associated with acoustic waves; rarefactions that follow shock-fronts lead to pressure decrease \citep[e.g.,][]{Kuzma2017,Kuzma2018} and consequently act contrary to plasma heating. 
Energy dissipated by these waves in the chromosphere is sufficient to heat up 
the plasma up to $10^4$ K on a time-scale of $10^3$~s. 
We conclude here that shorter waveperiod waves 
heat more lower atmospheric heights than higher waveperiod waves. In all cases, plasma heating grows with height, up to $y=1.5-1.8$~Mm. Higher up $\delta T_{\rm i}/T_{\rm i}$ falls off with height. 

%
%
\begin{figure}[!ht]
        \begin{center}
        \mbox{
                \includegraphics[scale=0.67, angle=0]{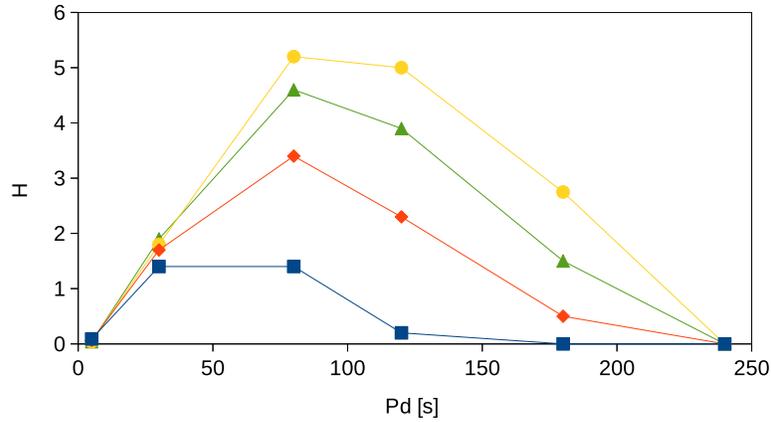}
                }
                     
                \caption{\small 
                Averaged over time relative perturbed temperature of ions, $H$, vs driving period $P_{\rm d}$, 
                at $y=1.0$~Mm (squares), $y=1.25$~Mm (diamonds), $y=1.5$~Mm (triangles) and $y=1.75$~Mm (circles). 
                        }
                \label{fig:4}
        \end{center}
\end{figure}

Figure \ref{fig:4} illustrates averaged over time relative perturbed temperature, $H$, 
given as
\begin{equation}
H(y) = \frac{1}{t_{a}} \int\limits_{0}^{t_{a}}\,\frac{T_{\rm i}(y,t)-T_{0}(y)}{T_{0}(y)}\, {\rm d {t}} \,  ,
\end{equation}
vs $P_{\rm d}$. Here ${t_{a}}=5\times 10^{3}$~s is the averaging time. 
%
%
The upper photosphere, at $y=0.5$ Mm, is a bit heated solely by short-period waves (not shown), 
while the chromospheric plasma is effectively heated by waves 
with their periods higher than $30$~s and shorter than 120 s (diamonds) with maximum close to $P_d=80$ s. Note that the upper chromosphere is heated by wider range of waves and more thermal energy is dissipated there (triangles and circles). 


The total vertical energy flux transported through the medium can be estimated as  
\begin{equation}
F(y,t) \approx \varrho_{i}(y) c_{{\rm s}}(y) V_{{\rm i} \, y}^{2}(y,t)+\varrho_{n}(y) c_{{\rm s}}(y) V_{{\rm n} \, y}^{2}(y,t) \, ,
\end{equation}
where 
$c_{\rm s}(y)$ 
is the sound speed at the equilibrium, given by 
\begin{equation}
\label{eq:soundspeed}
c_{\rm s}(y)=\sqrt{\frac{\gamma (p_{i}(y)+p_{n}(y))}{\varrho_{i}(y)+\varrho_{n}(y)}} \, .
\end{equation}
Wave energy is deposited in form of thermal energy to overlaying plasma by ion-neutral collisions. Figure 5 illustrates the collisional heating (first) part of the 
$Q^{\rm in}_{\rm i}$ term (see Eqs. 7-8) in the case of driving period $P_{\rm d}=80$~s. 
In Table 1 we compare $F(y, t=2.5 \cdot 10^3$ s), evaluated at $y=0.5$ Mm, $y=1.0$ Mm and $y=1.75$ Mm (from Fig. 3) that correspond to the lower, middle and upper chromosphere, with the corresponding radiative energy losses as estimated by \cite{Withbroe1977}. 
Note that numerically obtained values are lower than the predictions. 
Thus we conclude that acoustic waves with their realistic amplitudes in the chromosphere do not carry a sufficient amount of energy to compensate radiative losses. 

\begin{figure}[!ht] 
        \begin{center} 
      \hspace{-0.5cm}
        \mbox{
                \includegraphics[scale=0.55, angle=0]{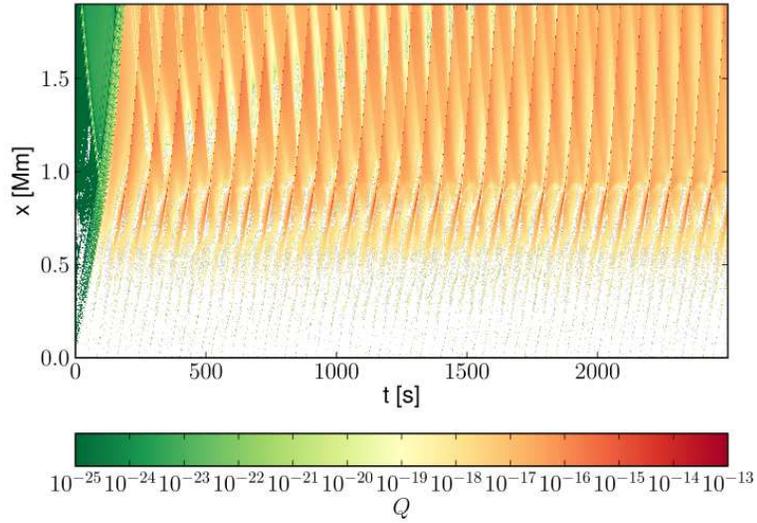} 
                }                   
\vspace{-0.5cm}
                \caption{\small Time-distance plot of the thermal energy heating rate due to ion-neutral collisions, taken from $Q^{\rm in}_{\rm i}$ term, 
                for $P_{\rm d}=80$~s, in units of W m$^{-3}$. 
                        } 
                \label{fig:6} 
        \end{center} 
\end{figure}

\begin{table}[h!] 
  \begin{center} 
    \caption{Radiative energy losses and averaged energy flux of acoustic waves in the solar chromosphere, $F(y, t=2.5 \cdot 10^3$ s), in the case of the driving period $P_{\rm d}=80$~s.}
    \label{tab:table1}
    \vspace{0.5cm}
    \begin{tabular}{|c|c|c|}
      \hline
      \textbf{} & \textbf{Radiative losses} & \textbf{F} \\ 
      {} & [erg cm$^{-2}$ s$^{-1}$]& [erg cm$^{-2}$ s$^{-1}$] \\
      \hline
      Low chromosphere & $2\cdot10^{6}$ & $2\cdot10^{3}$ \\
      Middle chromosphere & $2\cdot10^{6}$  &  $2\cdot10^{3}$ \\
      Upper chromosphere & $3\cdot10^5$ & $5\cdot10^{2}$\\
      \hline
    \end{tabular}
  \end{center}
\end{table}

We have verified by inspection that 
for 
waveperiods 
$P_{\rm d}=180$~s and $P_{\rm d}=300$~s 
consecutive heating shock-fronts are separated in space along $y$-direction. The localized plasma pressure increase is compensated by rarefactions which follow the leading shock-front, before the consecutive shock-front arrives to the same region, increasing the plasma temperature even more. Due to rarefactions plasma temperature attains its quasi-stationary value. 
As a result, the temporally averaged relative perturbed temperature, $H$, tends to constant (0 in case of $P_{\rm d}=300$~s) 
%
(Fig. 3, bottom-right, Fig. 4, bottom). 
Thus we infer that for very high period waves plasma heating does not accumulate in the chromosphere, 
and the oscillations 
with these characteristic periods cannot effectively heat the plasma. 




%
%
\section{Summary and conclusions}
We performed numerical simulations of two-fluid acoustic waves propagating in 
the gravitationally stratified and partially ionized solar photosphere and chromosphere. 
We perturbed the initial hydrostatic equilibrium with the periodic driver in 
vertical components of both ion and neutral velocities, 
which operates 
at the bottom of the photosphere. 
We found that waveperiods between about 30 and 200 s lead to significant heating of the chromosphere, 
while the upper photosphere remain hardly affected. 
The wave amplitude grows with height due to mass density falloff \citep[e.g.,][]{Murawski2018}, 
and thus the amplitude growth can dominate over ion-neutral collisions damping. 
Such waves possess the capability to penetrate the upper layers of the solar atmosphere 
and dissipate their energy in the chromosphere. 
However, for low-amplitude waves, the deposited energy remains low and plasma heating is insignificant there. 

By performing our two-fluid simulations, we contributed to ongoing discussions on plasma heating by means of acoustic waves. Previous works such as done by \cite{Carlsson2007}, \cite{Andic2008} and \cite{Sobotka2014} were based on observational estimations and single-fluid equations, and they reached the conclusion that energy carried by acoustic waves may compensate a substantial fraction of the radiative losses. Our results reveal that by implementing ion-neutral collisions, the plasma heating, which corresponds to the amplitude of the driver equal to 0.1 km s$^{-1}$ is not sufficient to compensate the radiative losses.


%
%
\section*{Acknowledgments}
The authors thank the referee for the stimulating comments, to Dr. Nikola Vinas for discussions on solar abundance, and to Dr. Istvan Ballai for stimulating discussions. 
This work was done in the framework of the projects from the Polish Science Center (NCN) grant No. 2014/15/B/ST9/00106, 2017/25/B/ST9/0050 and 2017/27/N/ST9/01798. The JOANNA code was developed by Darek W\'ojcik at UMCS, Lublin, Poland. 

\bibliographystyle{aasjournal}
\bibliography{draft.bib}

\end{document}